\documentclass{article}
\usepackage{spconf,amsmath,graphicx}
\usepackage{tabularx,booktabs}
\usepackage{lipsum}
\usepackage{multirow}
\usepackage{booktabs}
\usepackage{cite}

\title{Auditory-Based Data Augmentation for End-to-End Automatic Speech Recognition}
\name{Zehai Tu, Jack Deadman, Ning Ma, Jon Barker}
\address{University of Sheffield, Department of Computer Science, Sheffield, UK\\
\textit{\{ztu3, jdeadman1, n.ma, j.p.barker\}@sheffield.ac.uk}}

\renewenvironment{thebibliography}[1]{
  \begin{oldthebibliography}{#1}
    \setlength{\itemsep}{0.32em}
    \setlength{\parskip}{0em}
}
{
  \end{oldthebibliography}
}

\begin{document}
\ninept

\maketitle

\begin{abstract}
End-to-end models have achieved significant improvement on automatic speech recognition. One common method to improve performance of these models is expanding the data-space through data augmentation. Meanwhile, human auditory inspired front-ends have also demonstrated improvement for automatic speech recognisers. In this work, a well-verified auditory-based model, which can simulate various hearing abilities, is investigated for the purpose of data augmentation for end-to-end speech recognition. By introducing the auditory model into the data augmentation process, end-to-end systems are encouraged to ignore variation from the signal that cannot be heard and thereby focus on robust features for speech recognition. Two mechanisms in the auditory model, spectral smearing and loudness recruitment, are studied on the LibriSpeech dataset with a transformer-based end-to-end model. The results show that the proposed augmentation methods can bring statistically significant improvement on the performance of the state-of-the-art SpecAugment.
\end{abstract}


{\let\thefootnote\relax\footnote{{© 2022 IEEE. Personal use of this material is permitted. Permission from IEEE must be obtained for all other uses, in any current or future media, including reprinting/republishing this material for advertising or promotional purposes, creating new collective works, for resale or redistribution to servers or lists, or reuse of any copyrighted component of this work in other works.}}}

\vspace{-5mm}
\section{Introduction}
\label{sec:intro}

In recent years, significant progress has been made in automatic speech recognition (ASR) due to the success of deep learning. Deep learning-based end-to-end models, including Connectionist Temporal Classification (CTC) models~\cite{graves2006connectionist}, attention-based sequence-to-sequence (seq2seq) models~\cite{chorowski2014end}, and RNN-Transducers~\cite{graves2012sequence}, outperform conventional hybrid models~\cite{trentin2001survey} in various speech recognition tasks. However, the problem of overfitting when training large deep learning models remains a significant issue~\cite{caruana2001overfitting}. Regularization techniques are required. In addition to regularized optimisation approaches, \textit{data augmentation} can also be regarded as an effective data-space solution. 

\begin{figure}[thb]
  \centering
  \includegraphics[width=0.95\linewidth]{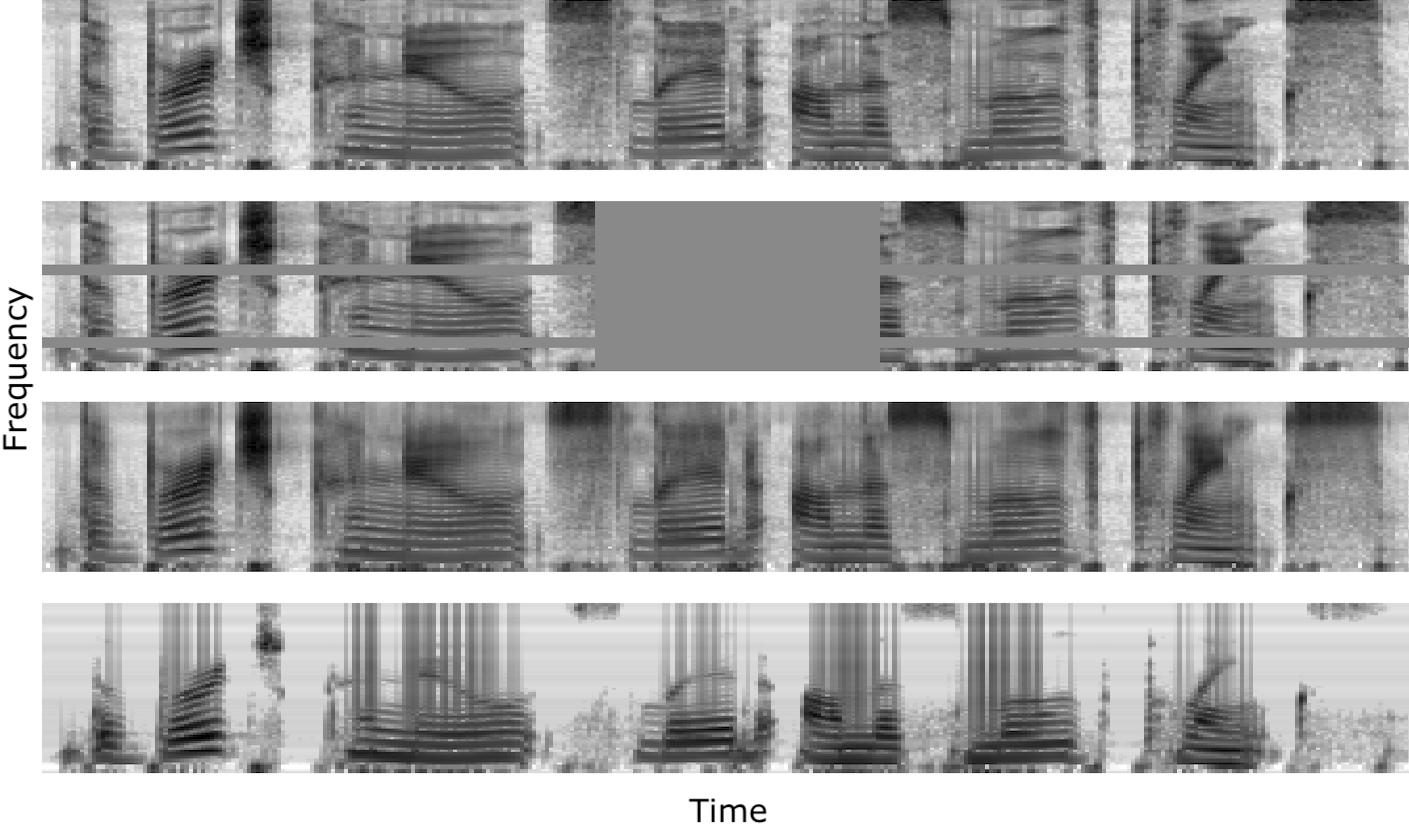}
  \caption{Augmentation methods applied to the time-frequency representation of a speech signal. From top to bottom, the figures show normalised features of the speech with no augmentation, SpecAugment (SA), spectral smearing (SS), loudness recruitment (LR).}
  \label{fig:specs}
  \vspace{-5mm}
\end{figure}


In contrast to most ASR systems, the human auditory system is capable of understanding speech even under poor conditions, including speech in noise and low-quality speech. One of the reasons is that listeners make use of many redundant cues in speech perception; if one set of cues becomes unavailable other strategies can be used. For example, there are studies that show that listeners with high-frequencies hearing loss start adopting different strategies for phoneme classification~\cite{varnet2019hearing}.

Leveraging auditory-based models has always been of interest to ASR researchers. In this work, we employ auditory-based models that simulate various degrees of hearing abilities as part of the data augmentation process. The underlying motivation is that such auditory models could introduce domain knowledge into end-to-end systems, by removing variation from the signal that cannot be heard and thereby encouraging the classifier to pay attention to other (more robust) consistent differences. 
Such transformations could reduce the chances of spurious correlations existing, which can be useful for wider domain generalisation, especially in noisy conditions where cues are lost due to noise masking.

This study exploits the Cambridge Auditory Group MSBG auditory model~\cite{baer1993effects, baer1994effects, moore1993simulation, stone1999tolerable} using a recent open-source implementation~\cite{graetzer21_interspeech}. Major auditory phenomena caused by defective hearing abilities, including reduced frequency selectivity and compressed loudness perceptual range, are modelled by spectral smearing (SS) and loudness recruitment (LR), respectively. The behaviours of these two mechanisms are dependent on individual differences in hearing ability. The training speech dataset is augmented with SS and LR by randomly sampling hearing ability related parameters. The augmented training set is then used to fine-tune a transformer-based ASR system that already incorporates the state-of-the-art SpecAugment (SA) data augmentation method.
An example time-frequency representation processed by SA, SS, and LR is shown in Fig.~\ref{fig:specs}.

This paper is organised as follows. Section~\ref{sec:relatedwork} reviews related work on auditory-based ASR and data augmentation for end-to-end systems. Section~\ref{sec:method} describes the proposed data augmentation methods and the end-to-end ASR models used in this study. Section~\ref{sec:experiments} presents the database and the experimental setup. The results comparing the performances of the SA baseline with the proposed SS or LR augmentation are presented and discussed in Section~\ref{sec:results}. Section~\ref{sec:conclusions} concludes the paper and presents ideas for future work.

\section{Related Work}
\label{sec:relatedwork}
Exploiting auditory inspired methods for ASR has been long studied. Two early examples are that the auditory perceptually based mel scale filterbank can improve recognition robustness especially in noise~\cite{jankowski1995comparison}, and the use of logarithm of speech energy features which approximates the nonlinear dynamic-range compression found in the auditory system. Additionally, a simulated auditory model based on findings from psycho-acoustical and physiological experiments was proposed as an ASR front-end and was found to improve speech-in-noise recognition~\cite{tchorz1999model}. Auditory inspired gammatone filterbanks~\cite{shao2009auditory, qi2013auditory} and Gabor filterbanks based on physiologically-motivated modulation frequencies~\cite{schadler2012spectro, meyer2012hooking} have also been used to improve ASR performance.  Recently, a number of works have presented auditory inspired methods for deep neural network based ASR~\cite{martinez2014should, baby2015investigating, martinez2017relevance}.

Various data augmentation methods have been proposed for ASR. Vocal tract length perturbation~\cite{jaitly2013vocal} was proposed to augment training utterances by randomising warp factors. Speech perturbation~\cite{kanda2013elastic, ko2015audio} and pitch adjustment~\cite{shahnawazuddin2016pitch} were proposed to adjust speed and pitch of the audio. Room impulse response simulation and adding point-source noises were proposed for far-field ASR~\cite{ko2017study}. Inspired by input dropout, \cite{toth2018perceptually} proposed to improve the noise robustness of CNN acoustic models by discarding input features, and \cite{kim2020small} proposed to mask time-frequency bins with energy lower than randomised thresholds. Recently, SpecAugment~\cite{park2019specaugment} (SA) was proposed to augment speech data by warping spectrograms along the time axis, and masking time and/or frequency bands in the spectral domain. Despite its simplicity, SA showed significant and consistent improvements for end-to-end speech recognition, and has become the standard approach for training state-of-the-art end-to-end ASR models. Methods similar to SA were also used for speech representation learning~\cite{ravanelli2020multi, kharitonov2021data}, and proved effective for downstream ASR tasks. Essentially, data augmentation encodes domain knowledge by disturbing the signal in ways `known' to not change its meaning.

\begin{figure}[t]
  \centering
  \includegraphics[width=0.9\linewidth]{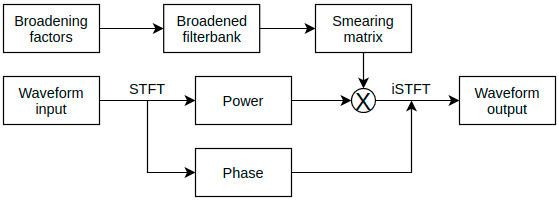}
  \caption{Procedure of the spectral smearing (SS) process.}
  \label{fig:smearing}
\end{figure}

\section{Method}
\label{sec:method}
In this section, the proposed data augmentation mechanisms SS and LR are described. To run the augmentation on the fly in the end-to-end framework for fast computation, the differentiable approximations~\cite{tu21b_interspeech} are used. Both SS and LR take the waveform speech signals as input, and generate waveform variants by randomly sampling hearing ability related parameters. The section concludes with a description of the end-to-end ASR model used for evaluation.

\subsection{Spectral smearing}
SS smooths the speech spectrum and suppresses details along the frequency axis by broadening the bandwidths of the auditory filters. This is similar to blurring for image data augmentation~\cite{shorten2019survey}.
In the MSBG model, SS is used to simulate the reduced frequency selectivity in the auditory systems of hearing impaired listeners~\cite{baer1993effects}. 
Subjective experiments showed that the effect on the intelligibility of spectrally smeared speech was minimal in quiet environments, while the affect would increase in noisy environments as signal-to-noise ratio drops~\cite{baer1993effects, baer1994effects}. Speech processed by SS can still preserve key recognition information if not heavily corrupted by noise.

The procedure for SS is shown in Fig~\ref{fig:smearing}. An input waveform is first converted into the time-frequency domain with an STFT. The power spectrogram is multiplied with the smearing matrix $A$, and the phase spectrogram is kept unchanged for later combination with the smeared power spectrogram. 

The smearing matrix $A_{S}$ is the product of the inverse of the normal auditory matrix $A_{N}$ and broadened auditory matrix $A_{W}$, i.e., $A_{S} = A_{N}^{-1}A_{W}$. An auditory matrix $A$ consists of a group of auditory filters, each of which has the the form of roex$(p)$:
\begin{equation}
    W(g) = (1 + pg)\exp(-pg),
\end{equation}
where $W(g)$ is the intensity weighting function describing the filter shape, $g$ is deviation from the centre frequency $f_{c}$ divided by $f_{c}$, and $p$ is the sharpness of the factor, which consists of lower side $p_{l}$ and upper side $p_{u}$ and where $p$ is computed as
\begin{equation}
        p = \frac{4f_{c}}{24.7\times(0.00437f_{c}+1)r}.
\end{equation}
$p_{l}$ and $p_{u}$ are computed with lower broadening factor $r_{l}$ and upper broadening factor $r_{u}$. Finally, each auditory filter is calibrated by being divided by $ (0.00437f_{c}+1)(r_{l} + r_{u})/2$. For normal hearing auditory filters, $r_{l}$ and $r_{r}$ are both $1$. For the purpose of data augmentation, random smearing matrices are generated by sampling $(r_{l}$, $r_{u})$ for broadened auditory filters---see Section~\ref{subsec:setup} for details.

\subsection{Loudness recruitment}

\begin{figure}[t]
  \centering
  \includegraphics[width=\linewidth]{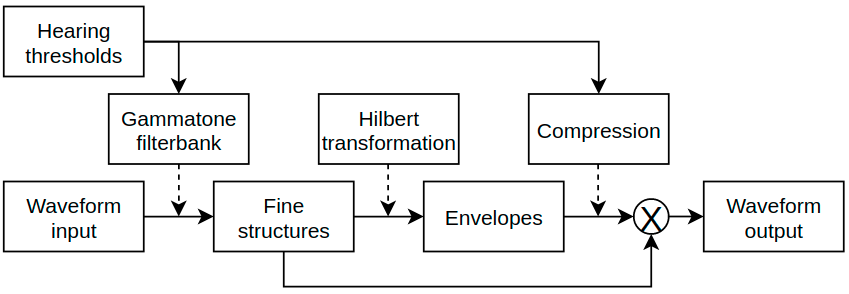}
  \caption{Procedure of the loudness recruitment (LR) process.}
  \label{fig:loudnessrecruitment}
\end{figure}

LR compresses amplitudes of different frequency bands according to the corresponding hearing losses. In the MSBG model, it is used to simulate the observation that the response of a damaged cochlea to low-level sounds is smaller than that of a healthy one, while the response to high-level sound is roughly the same~\cite{moore1993simulation}. Hearing impaired listeners can usually understand clean speech at adequate sound levels. By generating different hearing loss parameters, LR can apply random suppression to different frequency bands. 
Therefore, LR could introduce randomness to the relative levels of different frequency bands while conserving key recognition information.

The procedure for LR is shown in Fig.~\ref{fig:loudnessrecruitment}. Given the input waveform signal, a group of mel scale gammatone filters are first used to extract fine structure  $x(n)^{(i)}$ in the time domain. Each filter $h^{(i)}$ can be expressed as
\begin{equation}
    h^{(i)}(t) = A^{(i)} t^{\left(N^{(i)}-1\right)} e^{-2 \pi b^{(i)} t}\cos \left(2 \pi f^{(i)} t\right),
\end{equation}
where $f^{i}$ is the centre frequency, $b^{(i)} = 1.019 \times 24.7\times(0.00437f_{c}+1)$ is the bandwidth, $N$ is the order of gammatone filters and is set to 4 in this work. $A^{(i)}$ is the normalisation coefficient. The fine structure signals are then aligned and used to estimate the envelope $E^{(i)}$ via a Hilbert transformation. The envelopes are then smoothed with a low pass filter and used for compression. The waveform output $y$ is recruited as
\begin{equation}
    y(n) = \sum_{i=1}^{I}\left(\frac{E^{\left(i\right)}\left(n\right)}{E_{\theta}}\right)^{\left(\frac{\theta}{\theta - \text{HL}^{\left(i\right)}}-1\right)}x^{(i)}\left(n\right),
\end{equation}
where $\text{HL}^{(i)}$ is the hearing threshold at the centre frequency, $\theta$ is the maximal loudness threshold set as 105~dB, and $E^{\theta}$ is the corresponding envelope magnitude. A number of audiograms, i.e., the tables recording hearing thresholds at various frequencies, are sampled for the purpose of data augmentation, and details are described in Section~\ref{subsec:setup}.

\subsection{End-to-end ASR model}
Transformer architectures~\cite{vaswani2017attention} have been widely used for end-to-end ASR and have achieved impressive results.
The model used in this work consists of a transformer encoder, decoder, and a convolutional neural network (CNN) based front-end for better utilisation of global context~\cite{han2020contextnet}. Both the encoder and decoder consist of multiple transformer blocks, whose core component is the multi-head attention mechanism. 
For each time step, given the query $Q \in d_{k}$, key $K\in d_{k}$, and value $V\in d_{v}$ projected from the input features, the attention mechanism is expressed as
\begin{equation}
    \textit{Attention}\left(Q, K, V \right) = \textit{Softmax}\left(\frac{Q K^{T}}{\sqrt{d_{k}}} \right)V.
\end{equation}
The multi-head attention implements the attention mechanism $h$ times in parallel. For each attention computation, the projection matrices from the input features to the queries, keys, and values are different. The concatenation of $h$ attention heads is at last multiplied by a linear projection matrix. The encoder transformer blocks are built on a multi-head attention, with other operations including residual connections, layer normalisation, and a feedforward network. Based on the encoder transformer block structure, the decoder transformer blocks insert another multi-head attention over the output of the encoder stack, and apply masking so that the predictions only depend on previous decoded outputs.

The joint CTC-attention mechanism~\cite{kim2017joint}, which combines the idea of CTC~\cite{graves2006connectionist} and attention-based seq2seq~\cite{chorowski2014end} within the multi-task learning framework, is used for the optimisation. 
The core idea of CTC is to leverage intermediate label representation by allowing the repetition of labels and adding a special blank label. The CTC loss can be computed as
\begin{equation}
    \mathcal{L}_{CTC} = - \log \left( \sum_{\pi \in \beta^{-1}(l)} \prod_{m=1}^{M} P(z^{m}_{\pi_{m}}) \right),
\end{equation}
where $\beta$ is the many-to-one mapping function that removes the repeated intermediate labels and blank labels, $l$ is the target label sequence, $\pi_{m}$ is the intermediate label sequence including also blank labels, and $P(z^{m}_{\pi_{m}})$ is the probability of the observing network output intermediate label $\pi_{m}$ at time $m$. The seq2seq training scheme is used with the Kullback-Leibler divergence loss,
\begin{equation}
    \mathcal{L}_{KL} = \sum_{u} P(z_{u})(\log P(z_{u}) - \log P(\hat{z}_{u})),
\end{equation}
where $z_{u}$ and $\hat{z}_{u}$ are the $u$-th ground truth label and the predicted tokens, respectively. Label smoothing is applied for the seq2seq training. The CTC loss is computed on the encoder output, and the seq2seq uses the decoder output for loss computing. During training, the overall multi-task learning loss is then computed as
\begin{equation}
    \mathcal{L}_{MTL} = \lambda\mathcal{L}_{CTC} + (1 - \lambda)\mathcal{L}_{KL},
\end{equation}
where $\lambda$ is a predefined weighting coefficient.

\section{Experiments}
\label{sec:experiments}

\subsection{Database}
We evaluate our proposed augmentation method on the LibriSpeech database~\cite{panayotov2015librispeech}. All 960 hours of labeled utterances are used for training the ASR models. The word error rates (WER) on \textit{test-clean} and \textit{test-other} are reported. The difference between the \textit{test-clean} and \textit{test-other} sets is the quality of the utterances, with the quality of \textit{test-clean} being higher.

\begin{table}[t]
  \caption{Maximal $r_{l}$ and $r_{r}$ for SS of different hearing impairment degrees.}
  \vspace{.1cm}
  \label{tab:parasss}
  \centering
  \begin{tabular}{c|cc}
    \toprule
    &$r_{l}^{\max}$&$r_{u}^{\max}$\\
    \midrule
    Mild & 1.1 & 1.6 \\
    Moderate & 1.6 & 2.4 \\
    Severe & 2.0 & 4.0\\
    \bottomrule
  \end{tabular}
\end{table}

\begin{table}[t]
  \caption{Highest hearing thresholds at different frequencies for LR of different hearing impairment degrees.}
  \vspace{.1cm}
  \label{tab:paraslr}
  \centering
  \resizebox{\linewidth}{!}{
  \begin{tabular}{c|cccccc}    
  \toprule
    & 250\,Hz&  500\,Hz&  1\,kHz&  2\,kHz&  4\,kHz&  6\,kHz\\
    \midrule
    Mild & 10\,dB & 10\,dB & 10\,dB & 15\,dB & 30\,dB &40\,dB \\
    Moderate & 20\,dB & 20\,dB &25\,dB &35\,dB &45\,dB &50\,dB \\
    Severe & 55\,dB &55\,dB &55\,dB &65\,dB &75\,dB &80\,dB \\
    \bottomrule
  \end{tabular}
  }
\end{table}

\subsection{End-to-end ASR model}
The SpeechBrain~\cite{ravanelli2021speechbrain} LibriSpeech ASR transformer recipe is used to build the ASR model. 80-channel filterbank features are extracted as the input with a 25~ms window with a stride of 10~ms. The front-end context CNN consists of three 2D convolutional layers with kernel size of (3, 3, 1), stride size of (2, 2, 1), and out channel size of (128, 256, 512). The encoder and the decoder consist of 12 and 6 transformer blocks, respectively. For each transformer block, the number of attention heads is 8, and the size of the feedforward layer is 3072. The GELU activation function is used within the transformer blocks. The loss weighting coefficient $\lambda$ is set to 0.4.

To conserve computational resources, training of all models starts from the LibriSpeech model released by SpeechBrain\footnote{huggingface.co/speechbrain/asr-transformer-transformerlm-librispeech}, i.e., we finetune models for a further 15 epochs, rather than train from scratch. For the proposed methods, a half of the input signals within a batch are augmented during finetuning. For fair comparison, baseline performances are obtained with further finetuning using the same number of epochs but using only the baseline SpecAugment.

\subsection{Experimental setup}
\label{subsec:setup}
SA is applied as the data augmentation method for the baseline models. The released model from SpeechBrain is trained with SA, and SA is also applied when finetuning. The filterbank features are masked along both time and frequency axis with maximal two bins for each axis. The bin widths are sampled up to 30 and 40 for frequency axis and time axis, respectively. Time warping is applied with two-dimensional bicubic interpolation. 
To gain more confident results, the baseline experiments are repeated five times independently with different random seeds, i.e., five models are finetuned for 15 epochs from the SpeechBrain released model with SA only.

\begin{table}[t]
\caption{The WERs of the ASR models optimised with SA only, SA + SS, and SA + LR, with different degrees of hearing impairment (SA: SpecAugment, SS: spectral smearing, LR: loudness recruitment). Experiments of optimisation with SA only, and the best system (SA + LR with moderate hearing impairment) are run five times independently, and the standard errors of the WERs are shown.}
\vspace{.1cm}
\centering
\label{tab:results}
\begin{tabular}{@{}lr|ll@{}}
\toprule
\multicolumn{1}{l}{} & \multicolumn{1}{l|}{} & \multicolumn{1}{l}{\textit{test-clean}} & \multicolumn{1}{l}{\textit{test-other}} \\ \midrule
SA &                                         & 3.36\% \scriptsize{$\pm$ 0.01\%}       & 8.12\% \scriptsize{$\pm$ 0.03\%}       \\ \midrule \midrule
\multirow{3}{*}{SA + SS}    & mild           & 3.32\%                         & 8.07\%                         \\
                            & moderate       & 3.38\%                         & 8.05\%                         \\
                            & severe         & 3.32\%                         & 7.96\%                         \\ \midrule
\multirow{3}{*}{SA + LR}    & mild           & 3.27\%                         & 7.96\%                         \\
                            & moderate       & 3.28\%                         & 7.74\%                         \\
                            & severe         & 3.29\%                         & 7.83\%                         \\ \midrule \midrule
SA + LR                     & moderate       & \multicolumn{1}{l}{3.28\% \scriptsize{$\pm$ 0.01~\%}}       & \multicolumn{1}{l}{7.77\% \scriptsize{$\pm$ 0.03~\%}} \\ \bottomrule
\end{tabular}
\end{table}

To evaluate the proposed data augmentation methods, we finetuned the ASR models with SS and LR separately, both together with SA. Three parameter settings are used for both SS and LR simulating three hearing impairment degrees: mild, moderate, and severe. 

For SS, pairs of $r_{l} \in [1.001, r_{l}^{\max})$, and $r_{r} \in [r_{l}, r_{u}^{\max})$ are sampled. The values of $r_{l}^{\max}$ and $r_{u}^{\max}$ for different hearing impairment degree settings are shown in Table~\ref{tab:parasss}. For LR, the randomly sampled audiograms are represented by the hearing thresholds $\text{HL}$ at [250, 500, 1000, 2000, 4000, 6000]~Hz, and each threshold $\text{HL}^{f}$ at frequency $f$ is sampled in the range of $[\max(\text{HL}^{f'}|f' < f), \text{HL}^{f}_{max})$. The highest hearing thresholds $\text{HL}^{f}_{max}$ at these frequencies for different hearing impairment degrees are shown in Table~\ref{tab:paraslr}. The remaining parameters within SS and LR all follow the MSBG hearing impairment simulator. In the preliminary experiment, ASR models are optimised with the six proposed augmentation candidates, i.e., three for both SS and LR. And for the final results, the method with the overall best performance in the preliminary experiment is run five times and compared with the baselines.

\section{Results and Discussions}
\label{sec:results}
\begin{figure}[t]
  \label{fig:results}
  \centering
  \includegraphics[width=\linewidth]{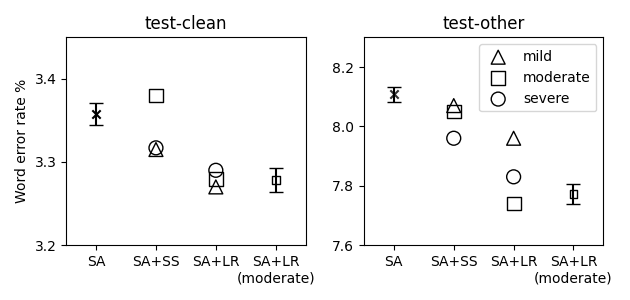} 
   \vspace*{-0.6cm}
  \caption{ 
A visual representation of the results presented in Table~\ref{tab:results}. Different settings of hearing impairment degrees are shown with different shapes. The results of optimisation with SA only, and SA + LR with moderate hearing impairment, are presented with mean WERs and error bars.}
  \label{fig:wers}
  \vspace*{-0.2cm}
\end{figure}

The results on LibriSpeech \textit{test-clean} and \textit{test-other} are shown in Table~\ref{tab:results} and visually presented in Figure~\ref{fig:results}. As SA is arguably the most successful data augmentation method for end-to-end ASR, it is used as the baseline. Results show that the proposed augmentation methods can further improve on the performance of SA.


The preliminary results show the optimisation with SA + SS using the mild and severe settings can achieve marginally lower WERs on \textit{test-clean} and \textit{test-other} compared to the average WERs of the baseline models, and the moderate setting performs worse than the baseline on \textit{test-clean}. We further probed the combination of SA + LR + SS, and found SS could not achieve lower WERs compared to SA + LR. In conclusion, SS could not bring improvement to the ASR models. On the other hand, the preliminary results show significant improvement can be achieved with LR augmentation. LR with the moderate hearing impairment setting performs best on \textit{test-other}, and very close to the best performance of \textit{test-clean} achieved by the mild setting. This could be related to the phenomenon that people with moderate hearing loss are able to understand clean speech, and severe hearing impairment setting could damage speech too much so that key information for recognition could be lost. 

Finally, repeated independent experiments of the optimisation of SA and LR with moderate hearing impairment with different random seeds are reported. SA can be further improved with the proposed method with relative 2.4\% and 4.3\% improvements on \textit{test-clean} and \textit{test-other}, respectively, and the improvements are statistically significant [$t$-test, $p<0.05$]. 

We further validate the performances with a transformer-based language model. The baseline models optimised with SA only can reach the WERs of 2.38\% \scriptsize $\pm$0.01\% \ninept and 5.50\% \scriptsize $\pm$0.02\% \ninept on the \textit{test-clean} and \textit{test-other}. Models optimised with SA + LR with moderate setting can achieve the WERs of 2.33\% \scriptsize $\pm$0.01\% \ninept and 5.32\% \scriptsize $\pm$0.03\% \ninept, with statistically significant relative improvements of 2.1\% and 3.3\% compared to the baseline [$t$-test, $p<0.05$].

\section{Conclusions}
\label{sec:conclusions}
In this work, two auditory mechanisms SS and LR from the well-verified MSBG hearing impairment model are studied for end-to-end ASR data augmentation. The mechanisms can augment speech datasets on the fly when optimising ASR models. The results show that LR with moderate hearing impairment setting can achieve statistically significant improvement on the performance of the state-of-the-art SA method on both \textit{test-clean} and \textit{test-other} of LibriSpeech when both using and not using an external language model. In the future work, the proposed data augmentation technique will be verified with more challenging speech recognition tasks, such as CHiME~5~\cite{barker2018fifth}.

\bibliographystyle{IEEEbib}
\bibliography{refs}

\end{document}